\documentclass[preprint,final,5p,times,twocolumn]{elsarticle}
\usepackage{rotating,color,subfigure,amssymb}
\usepackage{alphalph}
\usepackage{amsmath}
\usepackage[T1]{fontenc}

\usepackage{amssymb}

\journal{Physics Letters B}
\begin{document}

\begin{frontmatter}



\title{$\pi^{0}\pi^{0}$ Production in Proton-Proton Collisions at $T_{p}$=1.4 GeV}
\author[IKPUU]{The WASA-at-COSY Collaboration\\[2ex] P.~Adlarson}
\author[Erl]{C.~Adolph}
\author[ASWarsN]{W.~Augustyniak}
\author[PITue]{M.~Bashkanov}
\author[IPJ]{T.~Bednarski}
\author[MS]{F.S.~Bergmann}
\author[ASWarsH]{M.~Ber{\l}owski}
\author[IITB]{H.~Bhatt}
\author[HISKP]{K.--T.~Brinkmann}
\author[IKPJ,JCHP]{M.~B\"uscher}
% ...

\address[IKPUU]{Division of Nuclear Physics, Department of Physics and 
 Astronomy, Uppsala University, Box 516, 75120 Uppsala, Sweden}
\address[Erl]{Physikalisches Institut, Friedrich--Alexander--Universit\"at 
 Erlangen--N\"urnberg, Erwin--Rommel-Str.~1, 91058 Erlangen, Germany}
\address[ASWarsN]{Department of Nuclear Reactions, The Andrzej Soltan 
 Institute for Nuclear Studies, ul.\ Hoza~69, 00-681, Warsaw, Poland}
\address[PITue]{Physikalisches Institut, Eberhard--Karls--Universit\"at 
 T\"ubingen, Auf der Morgenstelle~14, 72076 T\"ubingen, Germany}
\address[IPJ]{Institute of Physics, Jagiellonian University, ul.\ Reymonta~4, 
 30-059 Krak\'{o}w, Poland}
\address[MS]{Institut f\"ur Kernphysik, Westf\"alische Wilhelms--Universit\"at
 M\"unster, Wilhelm--Klemm--Str.~9, 48149 M\"unster, Germany}
\address[ASWarsH]{High Energy Physics Department, The Andrzej Soltan Institute
 for Nuclear Studies, ul.\ Hoza~69, 00-681, Warsaw, Poland}
\address[IITB]{Department of Physics, Indian Institute of Technology Bombay, 
 Powai, Mumbai--400076, Maharashtra, India}
\address[HISKP]{Helmholtz--Institut f\"ur Strahlen-- und Kernphysik, 
 Rheinische Friedrich--Wilhelms--Universit\"at Bonn, Nu{\ss}allee~14--16, 
 53115 Bonn, Germany}
\address[IKPJ]{Institut f\"ur Kernphysik, Forschungszentrum J\"ulich, 52425 
 J\"ulich, Germany}
\address[JCHP]{J\"ulich Center for Hadron Physics, Forschungszentrum J\"ulich, 
 52425 J\"ulich, Germany}
% ...

\fntext[fnec]{present address: INFN, Laboratori Nazionali di Frascati, Via E. 
 Fermi~40, 00044 Frascati (Roma), Italy}
\fntext[fncp]{present address: Fachbereich Physik, Bergische Universit\"at 
 Wuppertal, Gau{\ss}str.~20, 42119 Wuppertal, Germany}

\cortext[coau]{Corresponding author }


\begin{abstract}
The reaction $pp$$\rightarrow$$pp$$\pi$$^{0}$$\pi$$^{0}$ has been investigated
at a beam energy of 1.4 GeV using the WASA-at-COSY facility. The total cross
section is found to be (324 $\pm$ $21_\text{systematic}$ $\pm$
$58_\text{normalization}$) $\mu$b. In order to study the production mechanism,
differential kinematical distributions have been evaluated. The differential
distributions indicate that both initial state protons are excited into
intermediate $\Delta(1232)$ resonances, each decaying into a proton and a
single pion, thereby producing the pion pair in the final state. No
significant contribution of the Roper resonance $N^{*}(1440)$ via its decay
into a proton and two pions is found. 

\end{abstract}



\end{frontmatter}


\section{Introduction}
\label{Int}
Investigations of the two-pion decay of mesons and baryons have been extensively carried out in pion-induced $\pi$$N$$\rightarrow$$\pi\pi$$N$ \cite{PhysRevC.69.045202} and photon-induced $\gamma$$N$$\rightarrow$$\pi\pi$$N$
\cite{Assafiri2003,Ahrens2005,Thoma:2007bm,SARANTSEV} reactions. Double pion production in nucleon-nucleon ($NN$) collisions is of particular interest in view of studying the simultaneous excitation of the two baryons and their subsequent decays. Here, the simplest case is considered: the excitation of the two nucleons into the $\Delta$(1232) resonance. {\bf The reaction reported on here provides the unique possibility to study this $\Delta\Delta$ process exclusively in very detail at its optimal energy of $T_p$ = 1.4 GeV, which corresponds to $\sqrt s$ = 2.48 GeV $\approx 2m_{\Delta}$.}
  
Several theoretical models for double pion production have been suggested in the energy range from the production threshold up to several GeV \cite{Tejedor1994667,Oset1985584}. A full reaction model describing the double pion production in $NN$ collisions has been developed recently by Alvarez-Ruso {\bf et al.(Valencia model)} \cite{AlvarezRuso1998519}. More {\bf recent} calculations by Cao, Zou and Xu include relativistic corrections {\bf not taken into account by the Valencia model, however, neglect interference between different reaction amplitudes} \cite{PhysRevC.81.065201}. These models include both resonant and non-resonant terms of $\pi\pi$-production {\bf and} predict {\bf the two-pion production process to be dominated by resonance excitation: At energies near threshold it is dominated by the excitation of one of the nucleons into the Roper resonance $N$$^{*}$(1440)$P$$_{11}$ via $\sigma$-exchange, followed by its s-wave decay $N$$^{*}$$\rightarrow$$N$($\pi$$\pi$)$^{\text{s-wave}}_{I=0}$ (where $I$ indicates the isospin of the $\pi\pi$ system). As the beam energy increases (i.e. $T_{p}$$>$1 GeV), the p-wave decay $N$$^{*}$$\rightarrow$$\Delta$(1232)$\pi$$\rightarrow$N($\pi$$\pi$) gives an increasingly growing contribution to the cross section. At higher energies ($T_{p}$$>$1.1 GeV) the double $\Delta$(1232) excitation is expected to become the dominant reaction mechanism.}
  
{\bf First measurements of two-pion production in $NN$-collisions stem from low-statistics bubble chamber measurements \cite{Shimizu1982571,PhysRev.138.B670}. More} recently, exclusive high-statistics measurements have become available from near threshold ($T_{p}$=650 MeV) up to $T_{p}$=1.3 GeV from the PROMICE/WASA \cite{Johanson200275,PhysRevLett.88.192301,PhysRevC.67.052202}, CELSIUS/WASA \cite{springerlink:10.1140/epja/i2008-10569-6,PhysRevLett.102.052301,Skorodko200930,Skorodko2011115,nnpipi}, COSY-TOF \cite{springerlink:10.1140/epja/i2008-10637-y}, {\bf WASA-at-COSY \cite {prl2011}} and {\bf COSY-}ANKE \cite{PhysRevLett.102.192301} experiments. The analysis of the data obtained from these experiments indicate that {\bf ,indeed as predicted,} in case of $pp$ collisions (isovector channel) only two $t$-channel reaction mechanisms dominate: the
excitation of the Roper resonance $N$$^{*}$(1440) at energies close to threshold \cite{PhysRevLett.88.192301,PhysRevLett.102.052301}, and the excitation of the $\Delta\Delta$ system at energies $T_{p}>$ 1.1 GeV \cite{Skorodko2011115}. {\bf In fact, the $pp$$\rightarrow$$pp$$\pi$$^{0}$$\pi$$^{0}$ reaction, which due to its isospin situation is the most suited reaction for studying these two resonance excitations \cite{Skorodko200930}, exhibits a distinctive dip in the slope of the total cross section separating the regions of dominance for Roper and $\Delta\Delta$   processes.}

Model predictions are found to be in good agreement with the experimental results at energies close to threshold, {\bf if the branching ratio for the decay $N$$^{*}$$\rightarrow$$\Delta$$\pi$$\rightarrow$N$\pi$$\pi$ is agjusted to the experimental findings \cite{SARANTSEV,PhysRevLett.88.192301,PhysRevC.67.052202,springerlink:10.1140/epja/i2008-10637-y,Skorodko2011115,tsEPJA}. At
energies $T$$_{p}$$\geq$1 GeV, the Roper resonance contribution to the total cross section is strongly over-predicted in the
Valencia calculations due to the too large branching ratio assumed there \cite{AlvarezRuso1998519}. As shown in Ref. \cite{Skorodko2011115} the Valencia calculation is also at variance with the differential data for the $\Delta\Delta$ process. However, if the $\rho$ exchange, which in the Valencia model is the dominating exchange process interfering destructively with the $\pi$ exchange, is strongly reduced and if also relativistic corrections are taken into account, then reasonable 
agreement with the data is obtained. Thus all three changes (modified Valencia model) lead then to a satifactory description of all data from threshold up to $T_p$=1.3 GeV \cite{Skorodko2011115}.

The most astonishing point in this result is that $\rho$ exchange obviously plays only a minor role in the $\Delta\Delta$ excitation. Though this agrees with the theoretical findings of Cao, Zou and Xu \cite{PhysRevC.81.065201}, naively one would have expected that the $\Delta\Delta$ process is a shorter-range phenomenon and hence is particularly sensitive to the $\rho$
exchange, since it involves already a considerable momentum-transfer. In order to study this result in more detail, it is desirable to investigate the $\Delta\Delta$ process at its optimal kinematic condition, which is reached at $\sqrt s = 2m_\Delta$ corresponding to $T_p\approx$1.4 GeV.} 
  
In contrast to the {\bf experimental situation} at energies $T_p$$\leq$1.3 GeV, there is little experimental information at higher energies. Only total cross sections are provided at $T_{p}$=1.36 GeV \cite{Koch2004} and $T_{p}$=1.48 GeV \cite{PhysRev.138.B670}.

Here, we report on {\bf exclusive and kinematically complete high-statistics } measurements of the $pp$$\rightarrow$$pp$$\pi$$^{0}$$\pi$$^{0}$ reaction at $T$$_{p}$=1.4 GeV using the WASA at COSY facility \cite{WASApro2004}. The beam energy corresponds to a center-of-mass energy of $\sqrt{s}$=2.48 GeV, $i.e.$ twice the $\Delta$ mass, thereby allowing a stringent test of the $t$-channel $\Delta\Delta$ mechanism. 

\section{Experimental Setup}
\label{Exp}
The experimental data were collected using the Wide Angle Shower Apparatus (WASA). WASA is an internal target experiment at the COoler SYnchrotron (COSY) of the Forschungszentrum J\"ulich, Germany. The detection system provides nearly full solid angle coverage for both charged and neutral particles. It allows multi-body final state hadronic interactions to be studied with high efficiency. The WASA facility consists of a central and a forward detector part and a cryogenic microsphere (pellet) target.
 
The pellet target generator is located above the central detector. It provides frozen pure hydrogen or deuteron pellets of about 25 $\mu$m diameter (as the targets), thereby minimizing background reactions from other materials.
 
The central detector is built around the interaction point and covers polar scattering angles between 20$^{\circ}$--169$^{\circ}$. The innermost detector, the mini drift chamber, is housed within the magnetic field of a
superconducting solenoid and is used in determining the momenta of charged particles. The next layer, the plastic scintillator barrel provides fast signal for first level trigger and charged particle identification. As the outermost layer, 1012 CsI(Na) crystals of the calorimeter enable the measurement of the energy deposited by charged particles as well as the reconstruction of electromagnetic showers. Due to the different size of the crystals, it was found that the energy and angle resolutions for the photons in the calorimeter are dependent on their energies and scattering angles, with average values of 15$\%$ and 1.5$^{\circ}$ for energy and angular resolutions, respectively.
 
The forward detection system covers the polar angular range of 3$^{\circ}$--18$^{\circ}$. The multi-plane straw tube detector is implemented for the precise reconstruction of charged particle track coordinates. An arrangement of segmented plastic scintillator layers, the forward range hodoscope, is used to reconstruct kinetic energies of scattered particles by the $\Delta$$E-E$ technique. A three-layered thin hodoscope provides fast charged particle discrimination. The forward detector can provide a tag on meson production via the missing mass of the reconstructed recoil particles. The proton energy resolution shows an approximately constant value of bout 3$\%$ for protons up to $T_{p}$=360 MeV which is the maximum energy for which protons can be stopped by the forward range hodoscope layers. The resolution worsens {\bf for more energetic protons and reaches about 20$\%$ for 1 GeV protons}. The angular resolution of the protons in the forward detector {\bf is} 0.15$^{\circ}$. The trigger for the present experiment demanded at least one charged particle candidate to reach the first layer of the forward range hodoscope. Figure 1 shows a schematic layout of te WASA detector at COSY, for more details about the WASA-at-COSY facility see Ref. \cite{WASApro2004}. 

\begin{figure}[hbt]
\centering
     
\includegraphics[width=.49\textwidth]{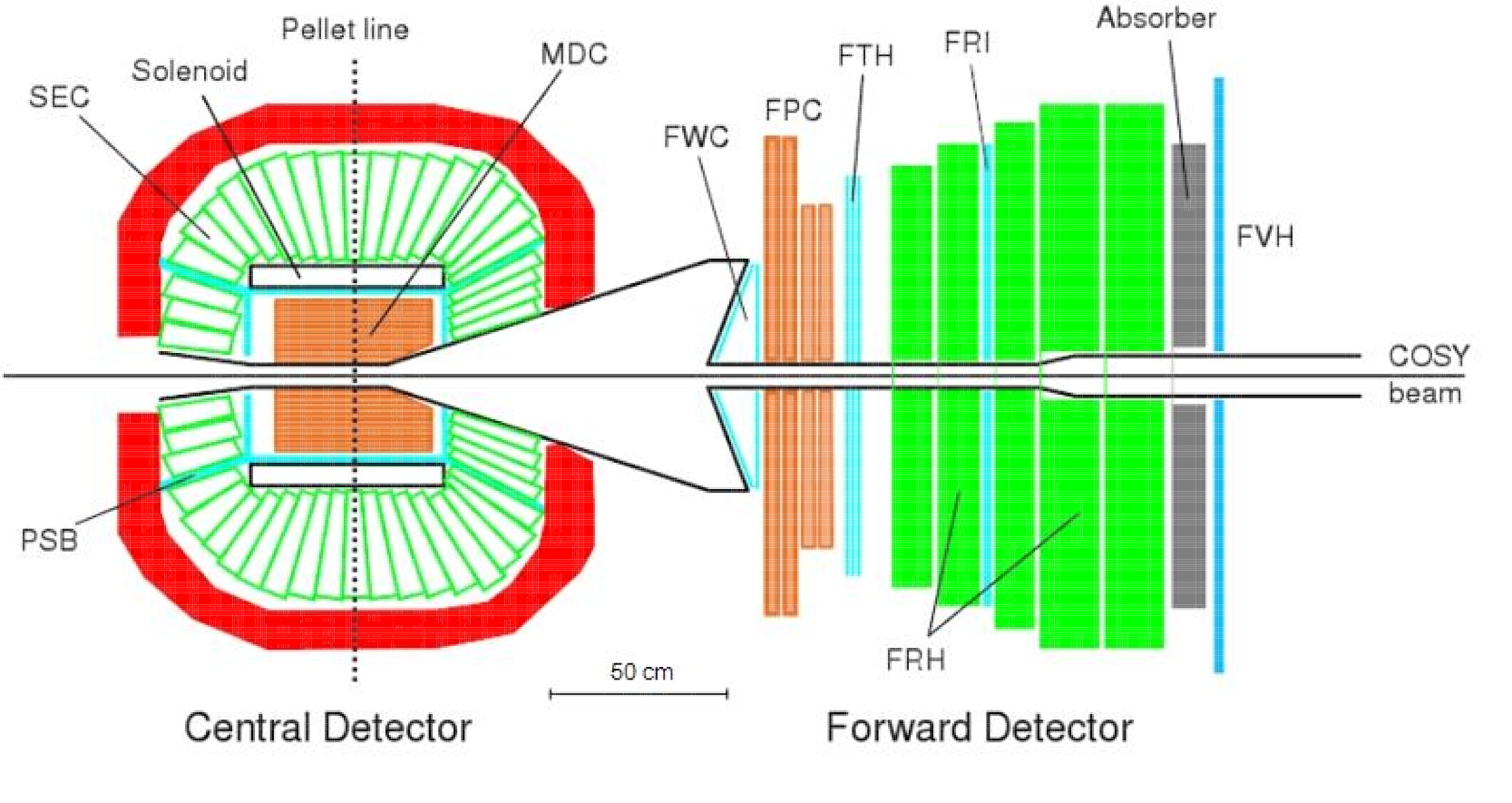}
\caption{The layout of the WASA detector at COSY. {\bf The SuperConducting Solenoid and the iron yoke for the return path of magnetic flux is shown shaded. Plastic scintillators are situated in the Plastic Scintillator Barrel (PSB), Forward Window Counters (FWC), Forward Trigger Hodoscope (FTH), Forward Range Hodoscope (FRH), Forward Range Intermediate Hodoscope (FRI), Forward Veto Hodoscope (FVH). Cesium Iodide scintillators are situated in the Scintillator Electromagnetic Calorimeter (SEC). Proportional wire drift tubes, straws, make up the Mini Drift Chamber (MDC) and the Forward Proportional Chambers (FPC).}
}
\end{figure}

\section{Data Analysis}
\label{DSel}
Recoil protons from the $pp$$\rightarrow$$pp$$\pi$$^{0}$$\pi$$^{0}$ reaction with $T_{p}$=1.4 GeV are detected in the forward detector, while the two neutral pions are reconstructed in the central detector. The main criterion to select the event sample demands 1 or 2 charged tracks in the forward detector and exactly 4 neutral tracks in the central detector. With this selection, the geometrical acceptance of the $pp$$\rightarrow$$pp$$\pi$$^{0}$$\pi$$^{0}$ reaction is found to be $45$$\%$. Two event samples are selected: the first includes events with only one proton detected in the forward detector while the other proton is scattered outside the forward detector. The second contains events when two protons were detected in the forward detector. The combination of both data samples gives a finite acceptance over all of the avaliable phase space, as shown in Fig. 2. Here, as an example, two two-dimensional acceptance distributions of $p\pi$$^{0}$ pairs (left plot) and of $p\pi^{0}\pi^{0}$- versus $p\pi$$^{0}$-invariant masses (right plot) are presented. The Monte Carlo plots are based on equally populated phase space and show that nearly the full phase space is covered. 
 
\begin{figure}[hbt]
\centering
     
\includegraphics[width=.23\textwidth]{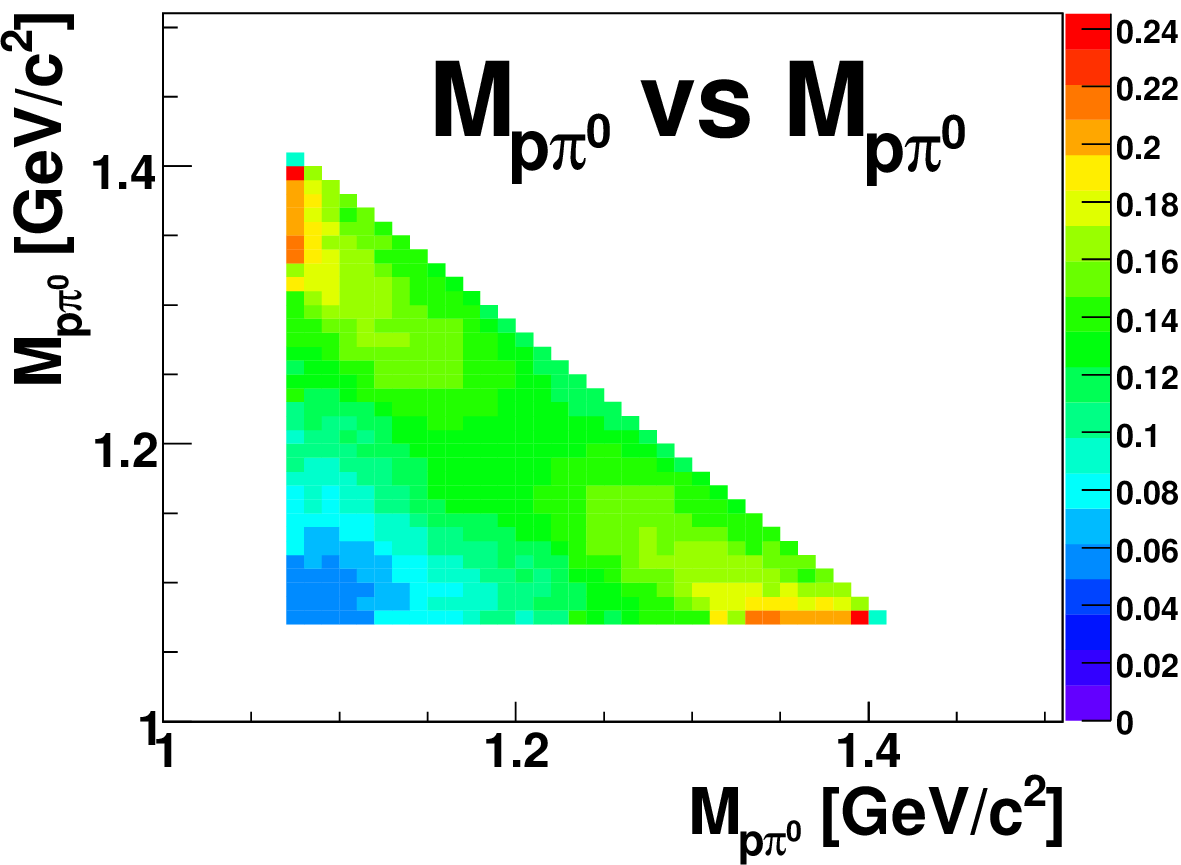}
      \includegraphics[width=.23\textwidth]{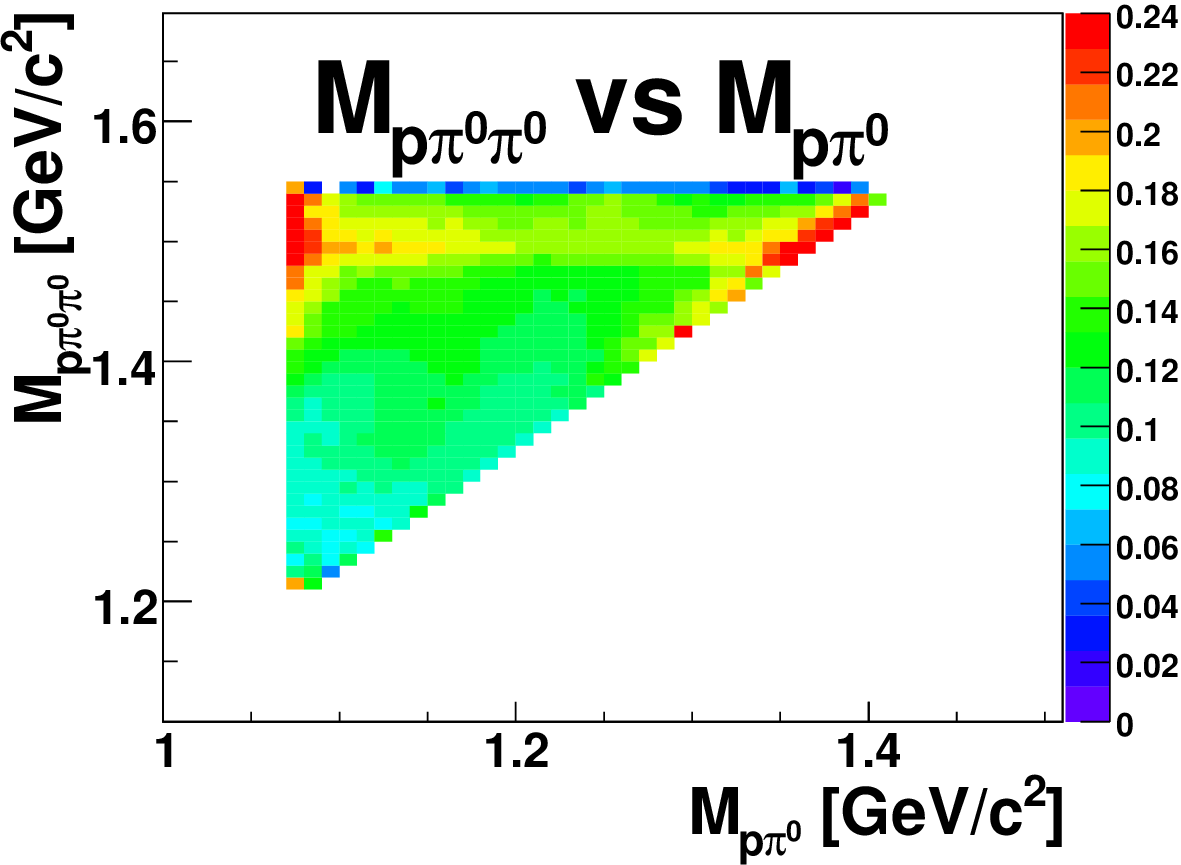}
      \caption{Product of the geometrical acceptance times the detector efficiency plotted as two-dimensional function of the $p\pi$$^{0}$-invariant mass pairs (left) and the invariant mass of the $p\pi^{0}\pi^{0}$ versus $p\pi$$^{0}$ (right).} 
\end{figure}

\begin{figure}[hbt]
\centering
     
\includegraphics[width=.235\textwidth]{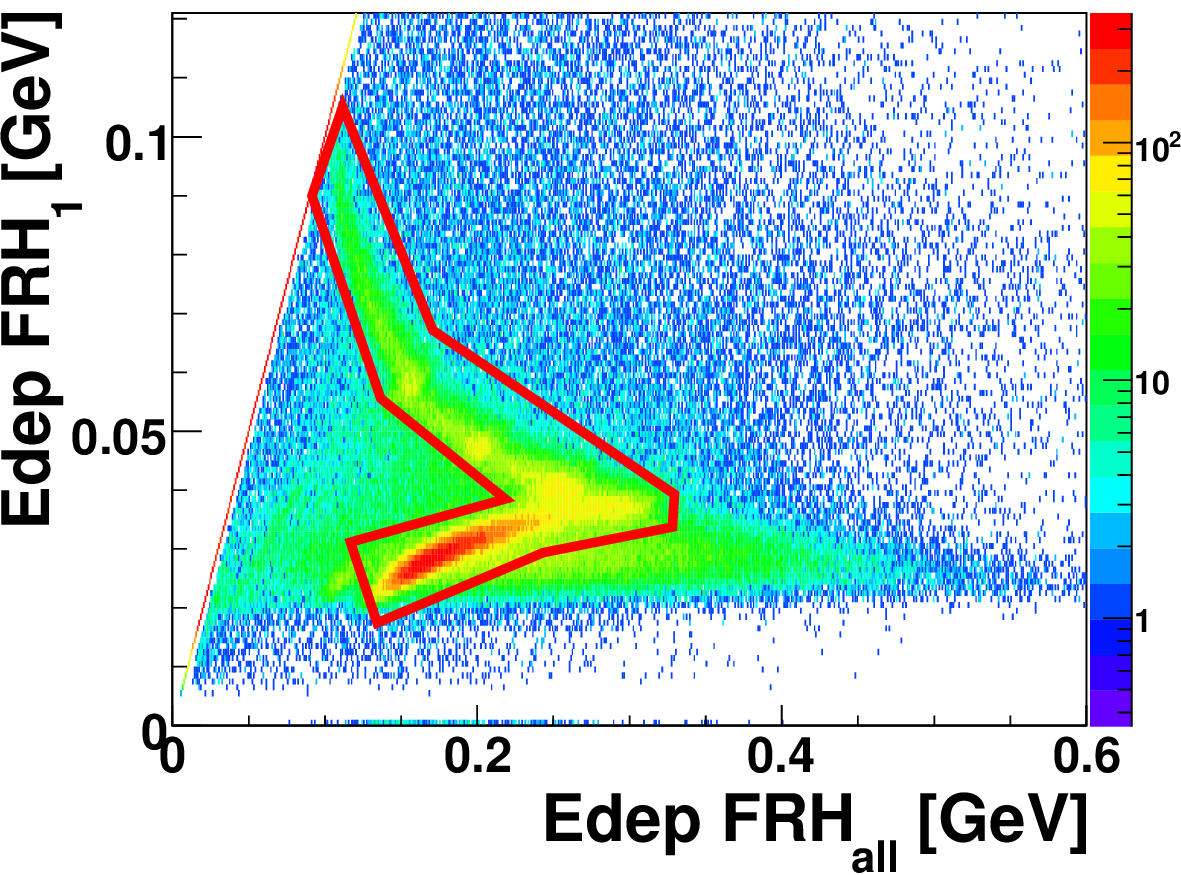}
 \includegraphics[width=.23\textwidth]{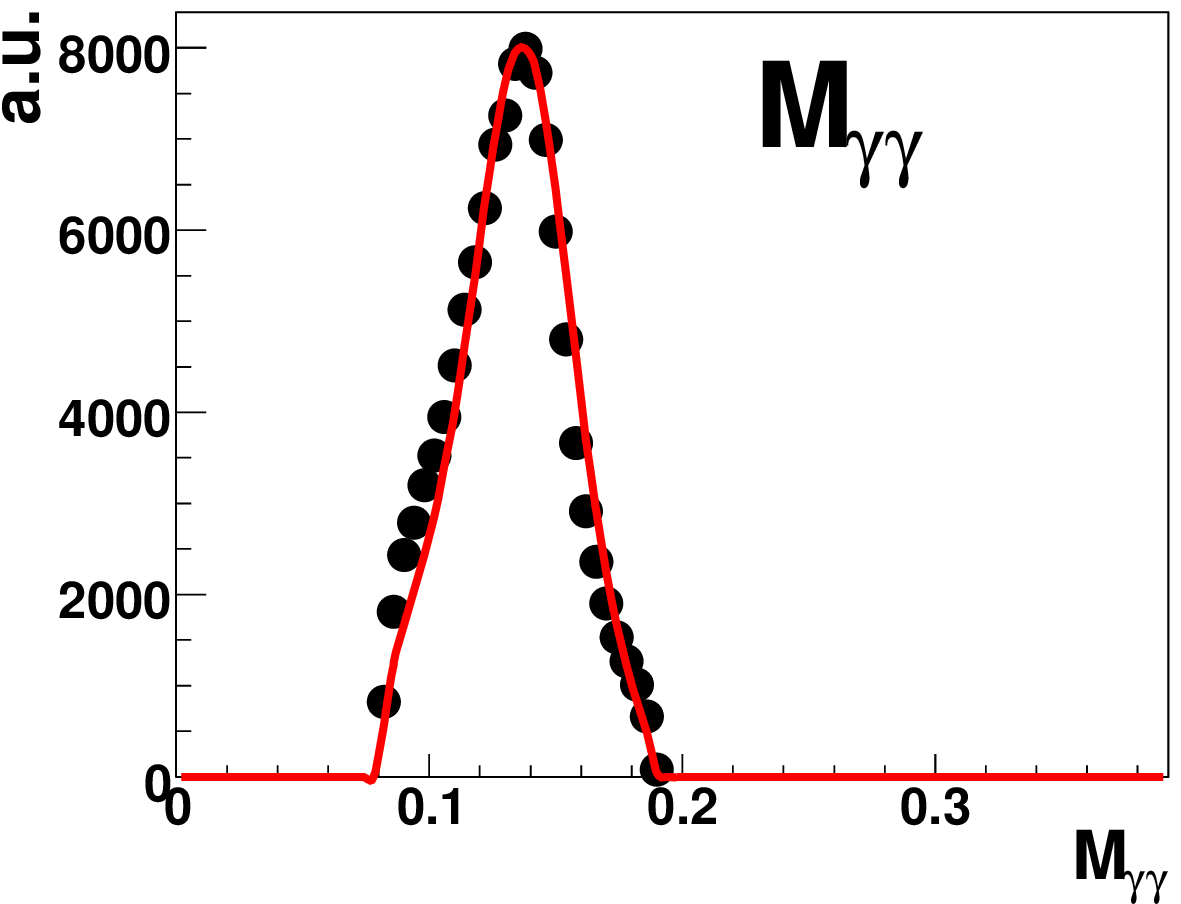}

      \caption{Left: $\Delta E-E$ {\bf spectrum} used for the identification of protons in the forward detector. Here, the deposited energy in the first layer of the forward range hodoscope (FRH) is plotted versus the total deposited energy in all layers of the forward range hodoscope. The {\bf solid} lines represent the region used to select protons. Right: {\bf spectrum of} the $\gamma$-pair invariant mass {\bf $M_{\gamma\gamma}$}. The {\bf for each event} best combination of 4 photons forming 2$\pi$$^{0}$  ({\bf obtained by the} $\chi$$^{2}$ method) is {\bf used}. {\bf Full} dots represent data points {\bf , whereas the solid}  line represents the Monte Carlo simulation.} 
\end{figure}

The identification of protons in the forward detector is based on the $\Delta E-E$ method, where the difference between the energy deposited in all layers of the detector (represented here by the FRH) and the energy deposited in a specific detector layer (represented here by the first layer of the FRH) is plotted as a function of the energy deposited in all layers of the detector, as shown in Fig. 3 (left plot). On one hand this technique is a powerful tool in distinguishing between the different particle species that are stopped in the detector. On the other hand, it is also used in distinguishing between
particles stopped in the detector elements and those, {\bf which} punch through. The depicted selection criterion (the red lines) selects not only protons that are stopped in the forward range hodoscope but also those that punch through. The
selection helps to reject the contribution resulting from hadronic interactions in detector material. The kinetic energy of the protons is reconstructed by translating the summed deposited energy over all the forward detector layers, {\bf after they have been} corrected {\bf for} the energy losses in the dead material between the detector layers as well as the quenching effect in the plastic scintillator. For more details about the particles identification and energy reconstruction see Ref. \cite{Tolba2010}.
 
Neutral pions have been reconstructed from the photon pairs detected in the central detector. The reconstruction procedure is based on the minimum $\chi$$^{2}$ method which is applied to select the two-photon combinations with invariant masses closest to the $\pi$$^{0}$ mass. Figure 3, right, shows the distribution of invariant masses ($M_{\gamma\gamma}$) for the best combination of the 4 photons forming two $\gamma\gamma$ pairs. The figure shows good agreement between the data points (full dots) and the Monte Carlo simulations (solid line). The figure also shows that the $M_{\gamma\gamma}$ distribution peaks at the $\pi$$^{0}$ mass with {\bf a resolution of} $\sigma$ = 18 MeV.

Furthermore, a kinematic fit with six constraints, four for total energy-momentum conservation and two for each of the two $\gamma \gamma$ pair masses being equate to the $\pi$$^{0}$ mass, is applied in order to suppress the contribution from background channels and to recover the information of the unmeasured proton, scattered into the central detector or into inactive material. For consistency, the kinematic fit routine is always applied with one unmeasured proton in the final state, with this assumption the number of constraints reduced to three. Hence, in the case where two protons are registered in the forward detector only one proton is selected and the other one is ignored. The proton with the lower energy is found to have better resolution. Therefore, it is chosen as the measured value in the kinematic fit routine while the higher energy one is treated as the unmeasured variable. In order to suppress events that {\bf do} not satisfy the kinematic fit conditions a cut-off at the 10$\%$ confidence level was applied. This specific cut was chosen because it has the {\bf largest} product of combinatorial purity {\bf and} reconstructed efficiency, where both data and Monte Carlo simulations are in the plateau region -- for more details see Ref. \cite{Tolba2010}.

The absolute normalization of the data has been achieved by normalizing to the measured $pp$$\rightarrow$$pp$$\eta$ cross section \cite{Chiavassa1994270}. Two decay modes of the $\eta$ meson, $\eta$$\rightarrow$$3\pi$$^{0}$ and $\eta$$\rightarrow$$2\gamma$, were chosen because they have similar final state particles as the $pp$$\rightarrow$$pp$$\pi$$^{0}$$\pi$$^{0}$ reaction. These channels have an additional advantage that they are the dominant neutral decay modes of the $\eta$ meson \cite{PDG2010}.

The data are corrected for the detector efficiency and acceptance by a Monte Carlo simulation using a toy model tuned to match the data. The toy model {\bf accounts for the previous findings that the $t$-channel $\Delta\Delta$ mechanism is expected to be the dominant effect and} is constructed by generating a four-body final state phase space distribution of the $pp$$\rightarrow$$pp$$\pi$$^{0}$$\pi$$^{0}$ reaction, employing the GEANT phase space generator, based on the FOWL program
\cite{James77}. Then, the generated event weight is modified to describe the 2$\pi$$^{0}$ production mechanism according to the production of two $\Delta$(1232)P$_{33}$ resonances in the intermediate state, each decaying into $p\pi$$^{0}$. The partial wave amplitude that describes the decay of $\Delta$(1232) into $p\pi$-system {\bf has been taken from} Ref. \cite{Risser197368}. This amplitude together with correction terms for the measured proton and pion angular distributions in the center-of-mass system, as well as for the $M$$_{\pi^{0}\pi^{0}}$ and $M$$_{p\pi^{0}}$ distributions are multiplied by
the generated weights of each event. The Monte Carlo simulations are then compared with the data, and this step is repeated until the data and the simulations are in good agreement. The tuned toy model is explained in detail in Ref. \cite{Tolba2010}.

\section{Results}
\label{Res}
The total cross section of approximately 500 k events of the $pp$$\rightarrow$$pp$$\pi$$^{0}$$\pi$$^{0}$ reaction at $T$$_{p}$ = 1.4 GeV is determined to be $\sigma_\text{tot}$=(324 $\pm$ $21_\text{systematic}$ $\pm$ $58_\text{normalization}$) $\mu$b. The total cross section error is evaluated in terms of statistical and systematic uncertainties. The statistical error is
found to be $<$1$\%$ and thus negligible compared to the systimatic contribution. The systematic error is constructed from two terms, systematic effects and normalization. The systematic contribution is estimated by observing the variation of the results with different analysis constraints where the varied parameters are assumed to be independent of each other. The
systematic term is calculated from the following main contributions: 1) applying different selection regions to the flat part of the confidence level (probability) distribution of the kinematic fit, the contribution from this term is found to be 5$\%$, 2) the contribution from the correction for the detector acceptance generated by different Monte Carlo models (the tuned toy
model, the model of Ref. \cite{Skorodko2011115} and the equally populated phase space model) is found to be 4$\%$, and 3) constraining the reconstructed particles to satisfy the geometrical boundaries of the central and the forward detectors, the contribution from this term is found to be 1$\%$. The total error from the systematic term is the square root of the quadratic sum of the individual terms and found to be 6.5$\%$. The normalization term is constructed from two main components: 1) contribution from the $pp$$\rightarrow$$pp$$\eta$ analysis, found to be 14$\%$, and 2) the uncertainty of the cross section value in Ref. \cite{Chiavassa1994270} which is found to be 11$\%$. The total error contribution from the normalization
term is estimated to be 18$\%$.

Figure 4 compares the cross section from this work (solid circle) with the previous experimental data \cite{Shimizu1982571,PhysRev.138.B670,Johanson200275,Skorodko200930,Skorodko2011115,Koch2004} and to the theoretical expectations calculated in Refs. \cite{AlvarezRuso1998519,Skorodko2011115}. The data point {\bf from this work} is compatible with the {\bf previous} results \cite{Shimizu1982571,PhysRev.138.B670,Johanson200275,Skorodko200930,Skorodko2011115,Koch2004}
and corroborates the strongly rising trend of the cross section starting at $\sim$ 1170 MeV. As has been verified in Ref. \cite{Skorodko200930}, the trend of rising total cross sections from threshold up to $T_{p}$$\sim$1 GeV is due to the dominance of the Roper resonance. Above 1 GeV, it levels off and proceeds with only a slight increase up to $T_{p}$$\sim$1170 MeV. The rise in the cross section values at higher energies $T_{p}$$>$1170 MeV is associated with the $\Delta\Delta$ excitation{\bf, as demonstrated in} Ref. \cite{Skorodko2011115}.  

\begin{figure}[hbt]
\centering
     \includegraphics[width=.40\textwidth]{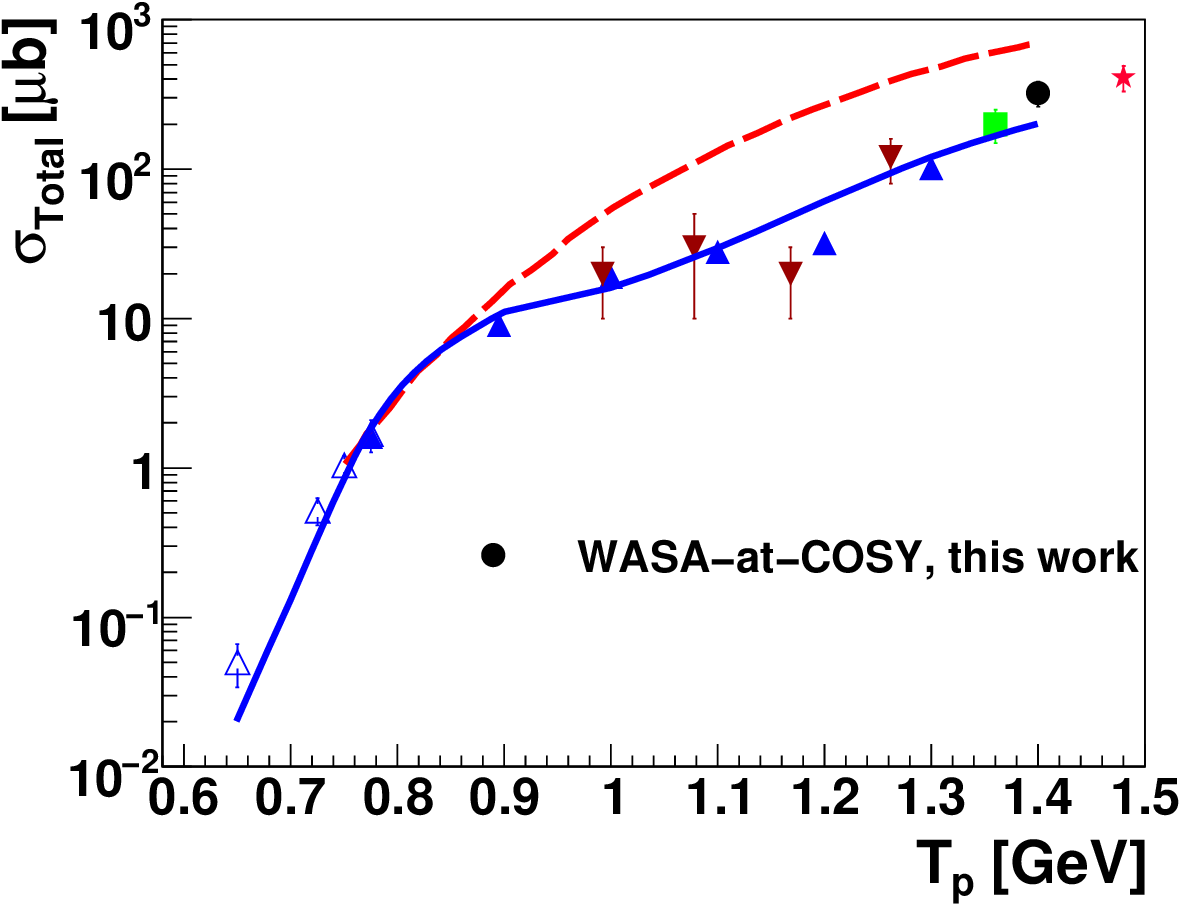}
     \caption{Total cross section for the $pp$$\rightarrow$$pp$$\pi$$^{0}$$\pi$$^{0}$ reaction as a function of $T_{p}$. The result of this work (solid circle), at $T$$_{p}$ = 1.4 GeV, is compared to the data from PROMICE/WASA (open triangles) \cite{Johanson200275}, CELSIUS/WASA (filled triangles) \cite{Skorodko200930} and at 1.36 GeV (square) \cite{Koch2004}, bubble chamber results (inverted triangles) \cite{Shimizu1982571} and (star) \cite{PhysRev.138.B670}, the theoretical calculations of Ref. \cite{AlvarezRuso1998519} ({\bf dashed line}) and of Ref. \cite{Skorodko2011115} ({\bf solid line), respectively}.}  
\end{figure}

In order to study the mechanism of the $pp$$\rightarrow$$pp$$\pi$$^{0}$$\pi$$^{0}$ reaction, seven independent kinematic variables are necessary to cover the available phase space of the reaction. Therefore, different kinematical variables describing the system have been investigated after the data have been corrected for the detector efficiency and acceptance using the tuned toy model. The corrected data are compared to an uniformly populated phase space distribution and the models
according to Refs. \cite{AlvarezRuso1998519} and \cite{Skorodko2011115}. All theoretical models are normalized to the same total cross section as the data. The differential distributions presented here have been chosen because they are sensitive to contributions from intermediate $\Delta$(1232) and/or the $N$$^{*}$(1440) resonances. 

\begin{figure}[hbt]
\centering
     \includegraphics[width=.23\textwidth]{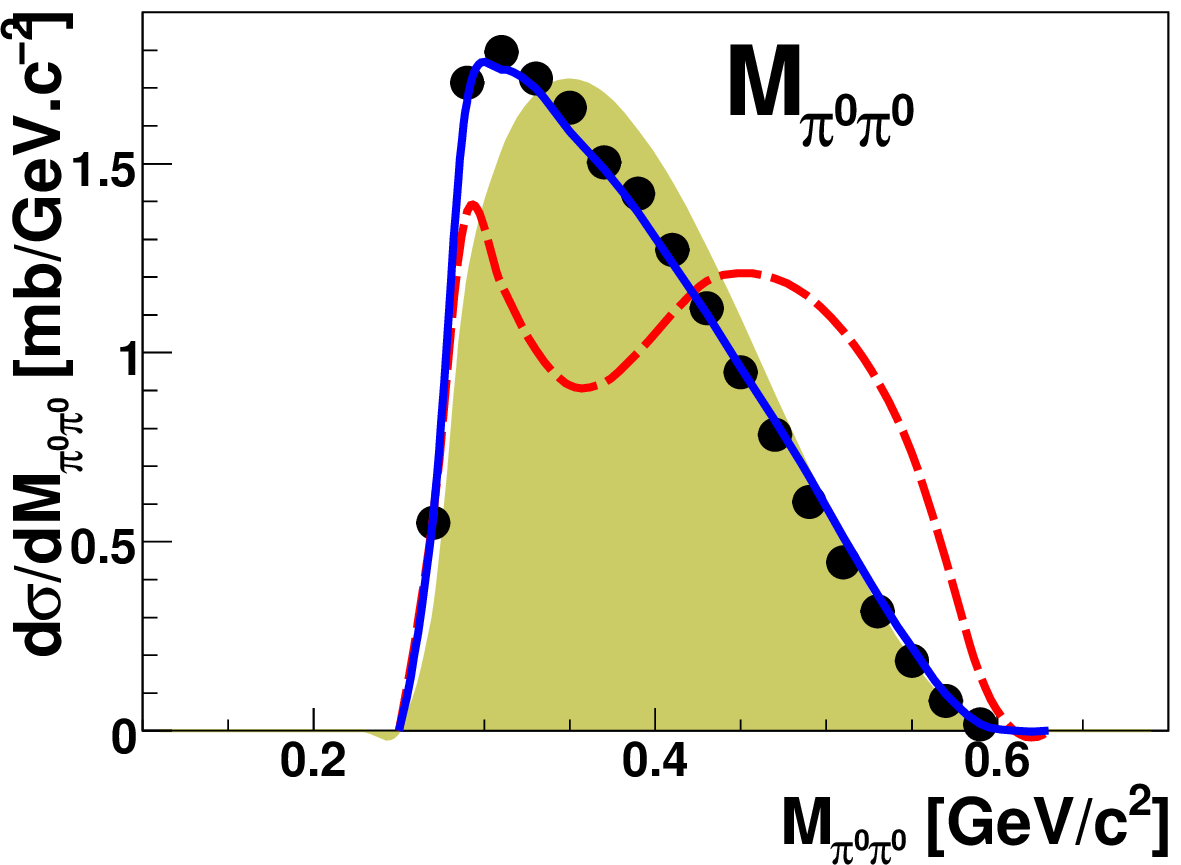}
     \includegraphics[width=.23\textwidth]{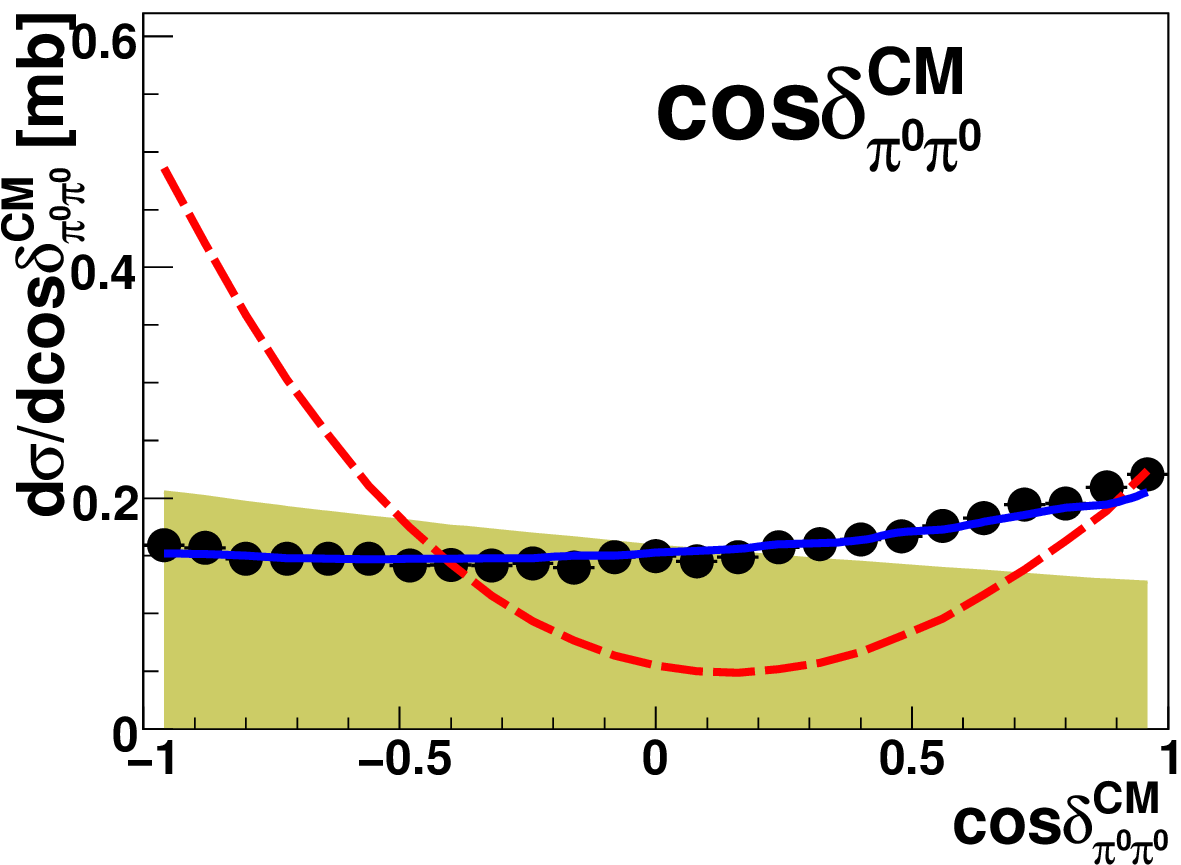}
     
    \caption{Comparison of data (black dots) to the theoretical expectations calculated from Ref. \cite{AlvarezRuso1998519} (red-dashed line) and Ref. \cite{Skorodko2011115} (blue-line), and with uniformly populated phase space (shaded area). Left: differential distribution of the $\pi$$^{0}$$\pi$$^{0}$-invariant mass, $M$$_{\pi^{0}\pi^{0}}$. Right: differential distribution of two pion opening angle in the center-of-mass system, $cos$$\delta$$^\text{CM}_{\pi^{0}\pi^{0}}$.} 
\end{figure}

Figure 5{\bf, left} shows that the $\pi$$^{0}$$\pi$$^{0}$-invariant mass ($M$$_{\pi^{0}\pi^{0}}$) distribution {\bf is closer} to the uniformly populated phase space distribution {\bf than to} the calculations of Ref. \cite{AlvarezRuso1998519}, {\bf which} predict two large enhancements at lower and higher $M$$_{\pi^{0}\pi^{0}}$ values. The enhancement at higher $M$$_{\pi^{0}\pi^{0}}$ values is due to the dominance of the $\rho$ exchange in the model calculations. In contrast, the
data are well described by the assumption of $t$-channel $\Delta\Delta$ excitation of Ref. \cite{Skorodko2011115} ({\bf solid} line), where the $\rho$-exchange contribution is strongly reduced compared to the original {\bf Valencia}calculations \cite{AlvarezRuso1998519}. The systematic enhancement at low $M$$_{\pi^{0}\pi^{0}}$ values indicates the tendency of the
two pions to be emitted {\bf in} parallel with respect to each other. This behavior is seen as well in the two pion opening angle distribution $cos$$\delta$$^\text{CM}_{\pi^{0}\pi^{0}}$ (right plot of Fig. 5), where the data is enhanced at $cos$$\delta$$^\text{CM}_{\pi^{0}\pi^{0}}$ = 1 relative to the phase space spectrum. Here, the data are well described by the modified calculations of Ref. \cite{Skorodko2011115}, {\bf whereas again} a large deviation is observed from the calculations of Ref. \cite{AlvarezRuso1998519}. The strong peaking of the latter calculations at an opening angle of 180$^{\circ}$ {\bf correlates with} the enhancement at higher values of $M$$_{\pi^{0}\pi^{0}}$ in left frame of Fig. 5. 

\begin{figure}[hbt]
\centering
     \centering
     \includegraphics[width=.23\textwidth]{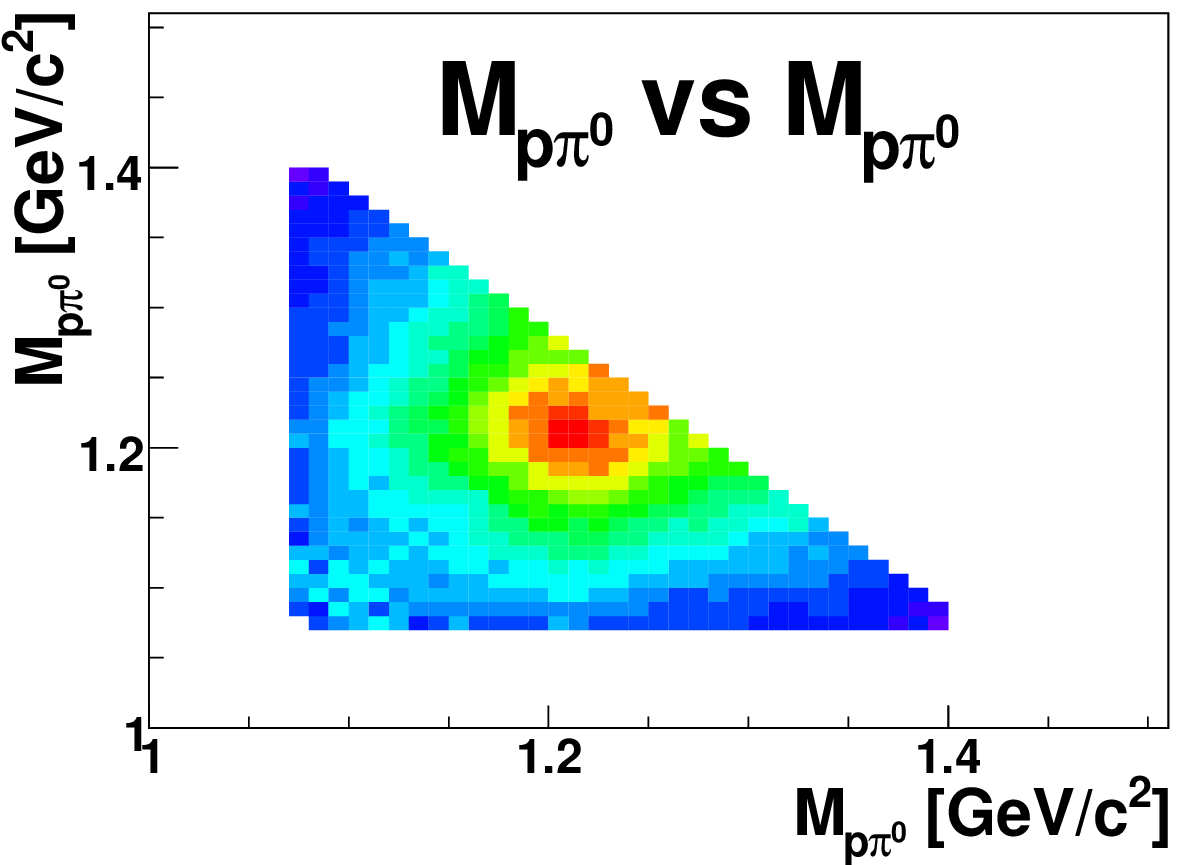} 
\includegraphics[width=.23\textwidth]{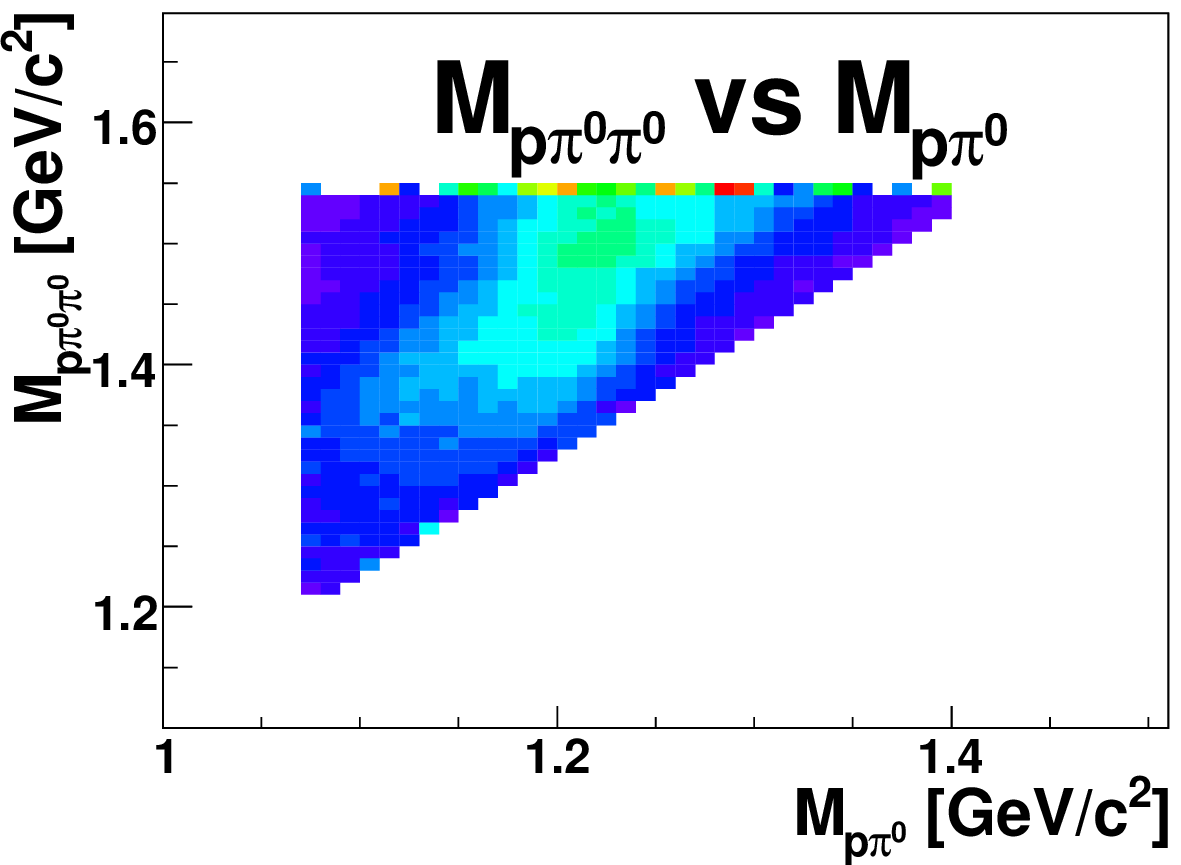}

     \centering
     \includegraphics[width=.23\textwidth]{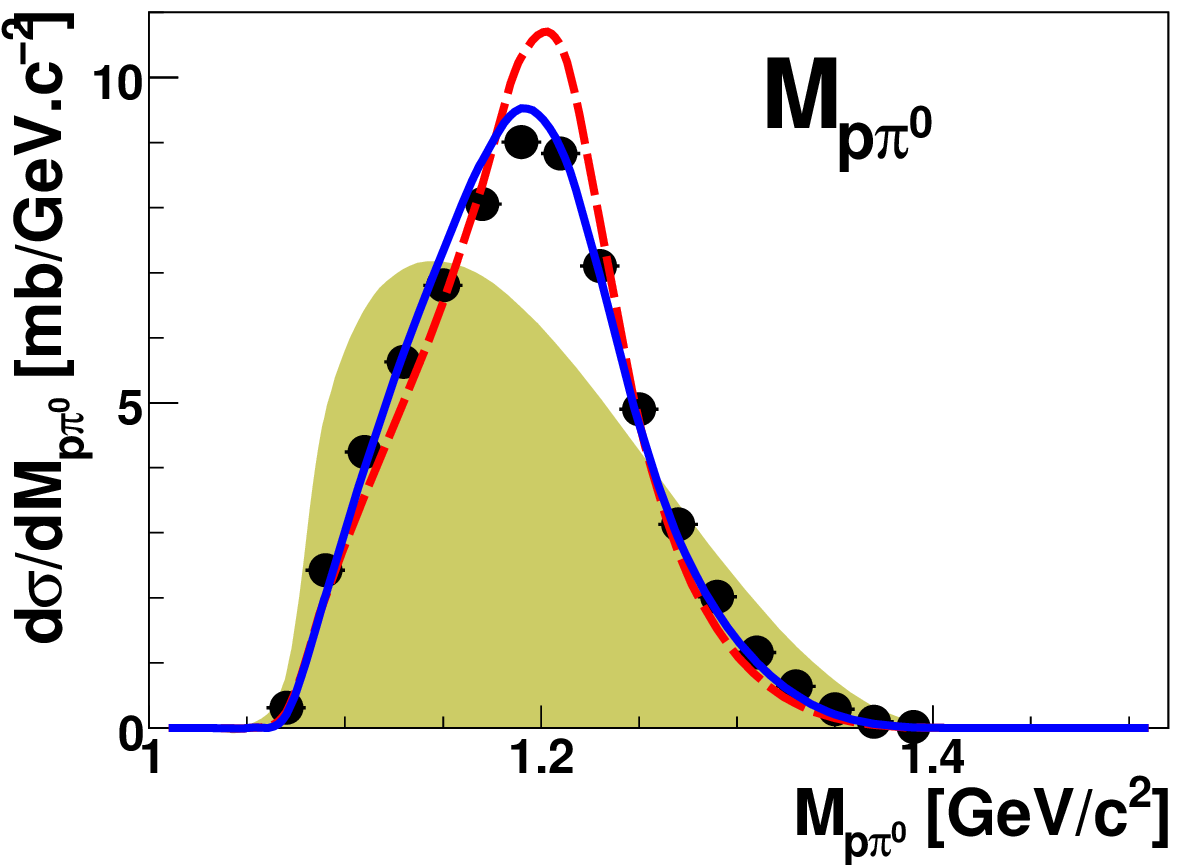} 
\includegraphics[width=.23\textwidth]{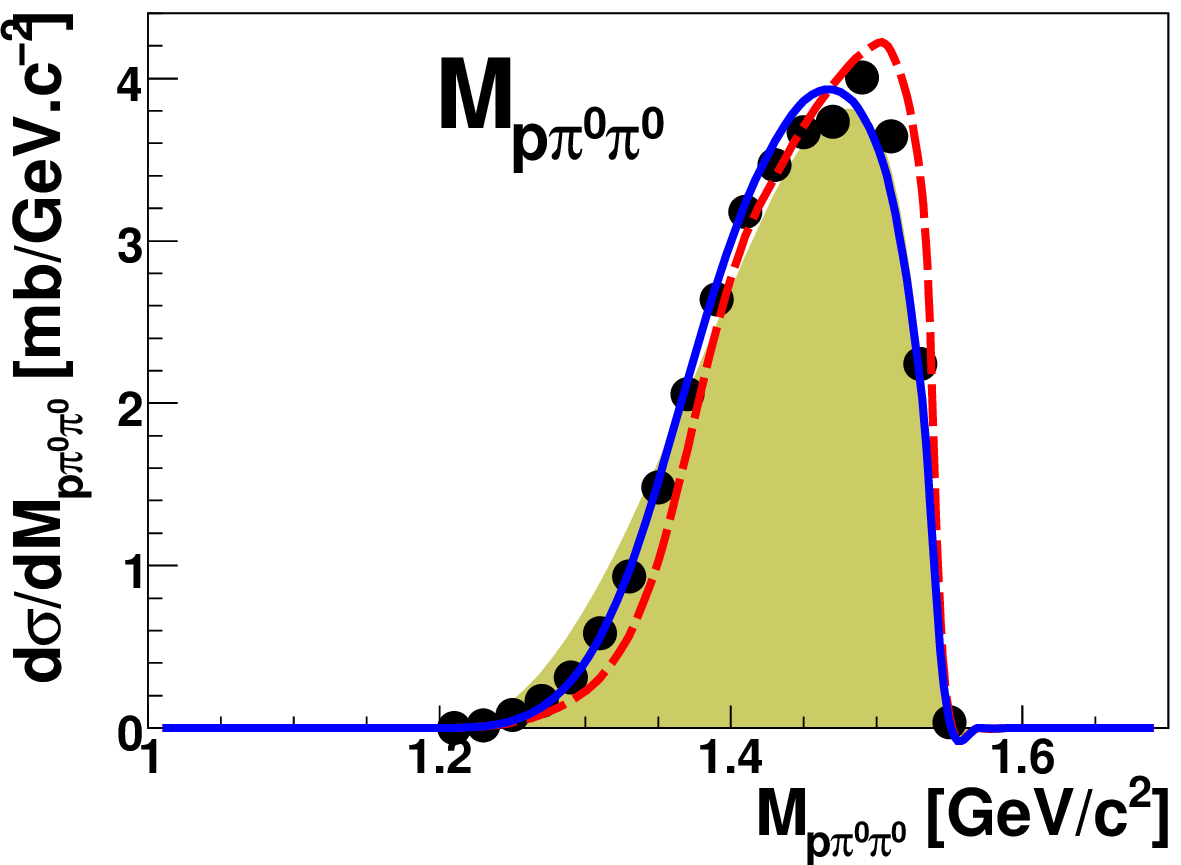}

     \caption{The upper frame shows two dimensional differential cross section distributions of $p\pi$$^{0}$- and $p\pi$$^{0}$$\pi$$^{0}$-invariant masses divided by the available phase space volume. This presentation enhances the sensitivity to resonance contributions in the production mechanism. The lower frame shows one-dimensional projections of the
differential cross section onto the $M$$_{p\pi^{0}}$ axis (left) and $M$$_{p\pi^{0}\pi^{0}}$ axis (right). See Fig. 5 for a description of the lines.} 
\end{figure}

The upper and lower left plots of Fig. 6 show indications for the $\Delta\Delta$ excitation in the correlation of the $M$$_{p\pi^{0}}$ pairs (upper) and in the one-dimensional projection onto the $M$$_{p\pi^{0}}$-axis (lower). Here, evidence for the $\Delta$(1232) resonance can be seen as a strong enhancement at $M$$_{p\pi^{0}}$ $\sim$ $M$$_{\Delta}$ = 1.232
GeV/c$^{2}$. The uniform phase space distribution of the lower plot shows the strongest deviation with respect to the data due to the $\Delta$ excitation in the data points. In contrast, no significant evidence for the presence of the Roper resonance $N^{*}$(1440) is observed in the right-hand plots of Fig. 6. Here, one would expect an enhancement around $M$$_{p\pi^{0}\pi^{0}}$ = 1.44 GeV/c$^{2}$. {\bf The small deviation of the data from the solid line at about 1.5 GeV could possibly signal some small contribution from the $N^*(1520)D_{13}$ resonance. However, since we are here close to the edge of the covered phase space, a solid statement on this matter can not be made.}
 
The upper left plot of Fig. 7 shows the $pp$-invariant mass spectrum, $M$$_{pp}$, which {\bf behaves complementary to the $M_{\pi^0\pi^0}$ spectrum in Fig. 5 and hence} peaks slightly to the right with respect to the uniformly populated phase space distribution. The upper right plot shows the angular distribution of the protons in the center-of-mass frame,
$cos$$\theta$$^\text{CM}_{p}$. It exhibits an anisotropic behavior, in agreement with the theoretical calculations. The strong forward-backward peaking of the $cos$$\theta$$^\text{CM}_{p}$ spectrum is associated with $\pi-\rho$ exchange mediating the $pp$ interaction. The angular distribution of the $p\pi$$^{0}$-system in the center-of-mass frame, cos$\theta$$^\text{CM}_{p\pi^{0}}$, (lower right) shows a forward-backward symmetry. That is similar in shape to the $cos$$\theta$$^\text{CM}_{p}$ distribution as expected from the large p/$\pi$$^{0}$ mass ratio. In the $pp\pi$$^{0}$-invariant mass distribution, $M$$_{pp\pi^{0}}$, (lower left plot) the data peak near the sum of the proton and $\Delta$ masses as expected for $\Delta\Delta$ production at threshold. Here, the modified model \cite{Skorodko2011115} and the data points are in good agreement, whereas the calculations of Ref. \cite{AlvarezRuso1998519} are shifted towards lower $M$$_{pp\pi^{0}}$ values. The data deviate strongly from the {\bf phase space distribution, but are again in favor of the $\Delta\Delta$ excitation process -- consistent with the observations in the other invariant mass distributions}.  

\begin{figure}[hbt]
  \centering
      \centering
      \includegraphics[width=.23\textwidth]{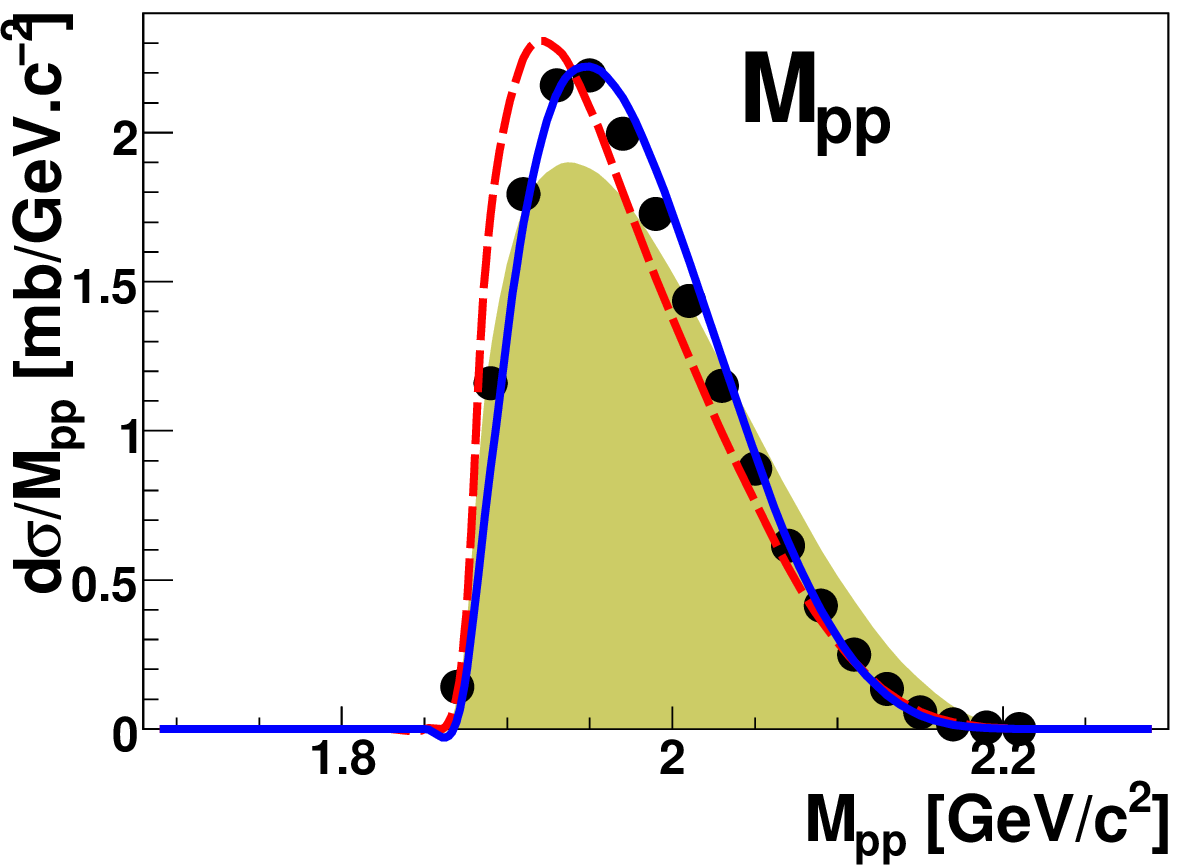}
      \includegraphics[width=.23\textwidth]{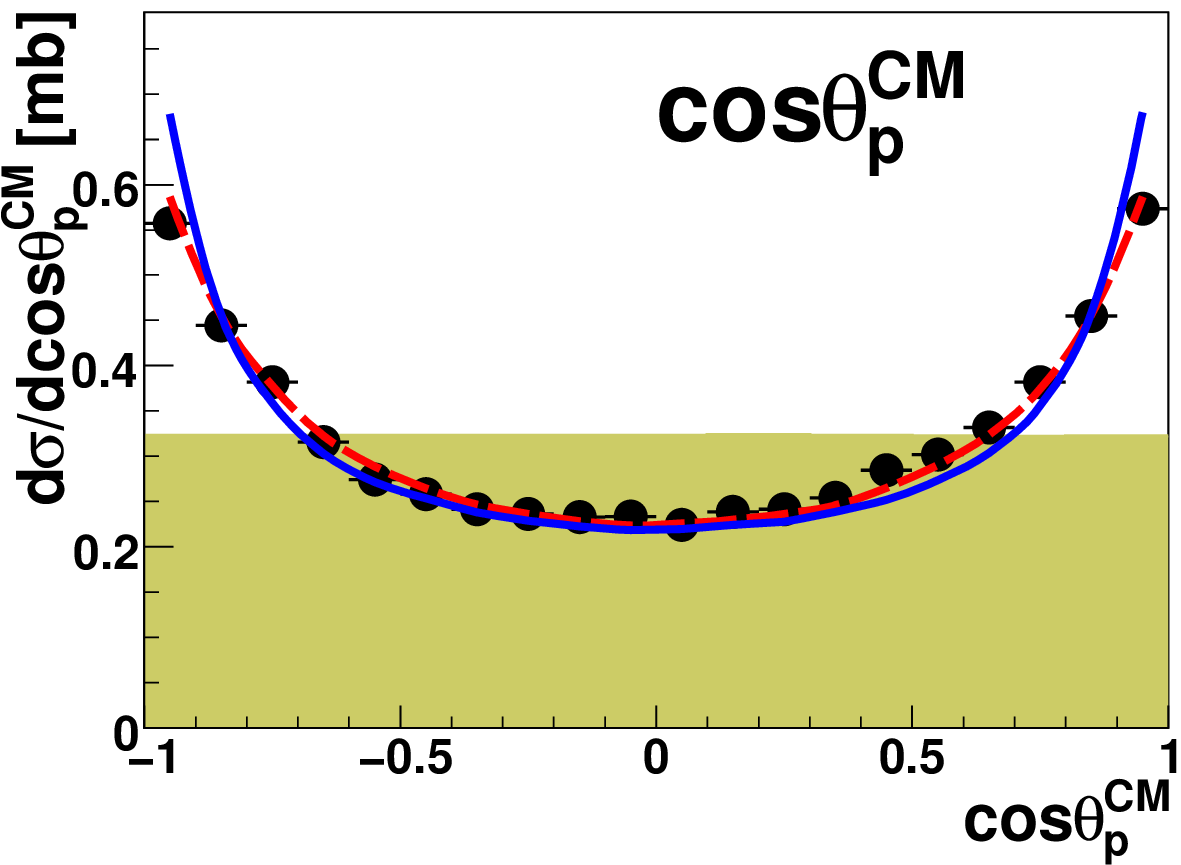}
      
      \centering
      \includegraphics[width=.23\textwidth]{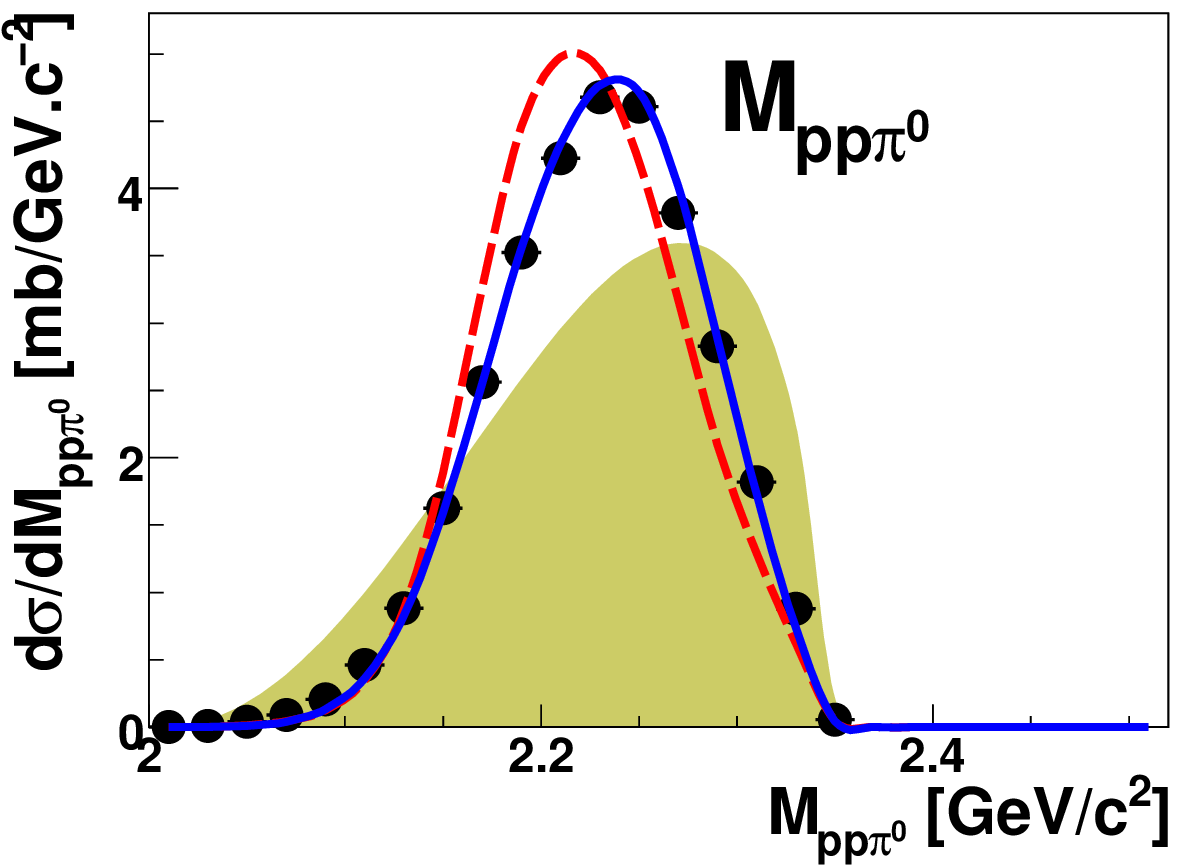}
      \includegraphics[width=.23\textwidth]{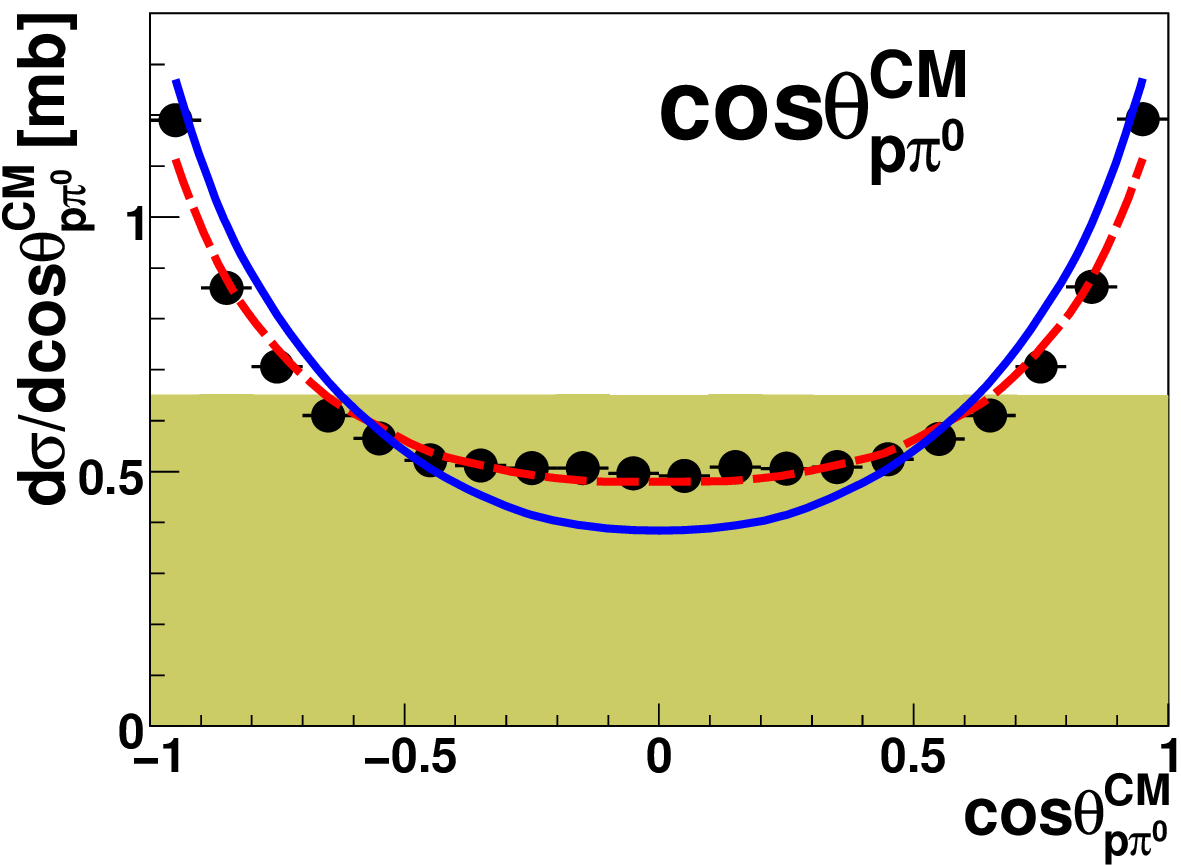}
      
  \caption{Same as Fig. 5 but for differential cross sections of $M$$_{pp}$ (upper left), $cos$$\theta$$^\text{CM}_{p}$ (upper right), $cos$$\theta$$^\text{CM}_{p\pi^{0}}$ (lower right) and $M$$_{pp\pi^{0}}$ (lower left).} 
  \end{figure}

\section{Conclusions and outlook}
\label{con}

{\bf The first exclusive and kinematically complete measurements at $T_p$=1.4 GeV reveal the $t$-channel $\Delta\Delta$ excitation to be the by far dominating process, whereas the Roper excitation is found to play no longer any significant role in the observables at such a high incident energy. The invarinat mass distributions are characterized by the $\Delta\Delta$
process, which meets an optimal condition with the incident energy corresponding to $\sqrt s$ = 2.48 GeV $\approx 2m_{\Delta}$. The modified Valencia model \cite{Skorodko200930} developed for the description of two-pion production at lower energies gives a good account for the new measurements reported here. The most astonishing conclusion from this good greement between data and calculations is that counter intuitively and in contrast to the original Valencia calculations \cite{AlvarezRuso1998519} the $\rho$ exchange does obviously not play a dominant role in the $t$-channel $\Delta\Delta$ process. 

The investigation of the production of charged pions ($\pi$$^{+}$$\pi$$^{-}$) is the next step in the study of the double pion production in $NN$ collisions with WASA-at-COSY. This channel is of special interest in order to study the {\bf production of} $\rho$$^{0}$(770), which is expected to play an important role in the $\pi$$^{+}$$\pi$$^{-}$ {\bf channel at higher energies}. Moreover, the extension to higher proton energies will shed light on the role of heavier resonances. 

\section{Acknowledgments}
\label{Ack}
This work was in part supported by: the Forschungszentrum J\"ulich including the COSY-FFE program, the European Community under the FP7-Infrastructure-2008-1, the German-BMBF, the German-Indian DAAD-DST exchange program, VIQCD and the German Research Foundation (DFG), the Wallenberg Foundation, the Swedish Research Council, the G\"oran Gustafsson Foundation, the Polish Ministry of Science and Higher Education and the Polish National Science Center (grant No. 0320/B/H03/2011/40) and Foundation for Polish Science - MPD program.
  
We also want to thank the technical and administration staff at the Forschungszentrum J\"ulich and at the participating institutes.
 
This work is part of the PhD thesis of Tamer Tolba.












\end{document}